\documentclass[twocolumn,showpacs]{revtex4}

\usepackage{graphicx}% Include figure files
\usepackage{dcolumn}% Align table columns on decimal point
\usepackage{bm}% bold math

\input epsf

\begin{document}

\title{\Large Entropy Bound of Horizons for Accelerating, Rotating and Charged Plebanski-Demianski Black Hole}

\author{\bf Ujjal Debnath\footnote{ujjaldebnath@yahoo.com ,
ujjal@iucaa.ernet.in}}

\affiliation{Department of Mathematics, Indian Institute of
Engineering Science and Technology, Shibpur, Howrah-711 103,
India.\\}

\date{\today}

\begin{abstract}
We first review the accelerating, rotating and charged
Plebanski-Demianski (PD) black hole, which includes the
Kerr-Newman rotating black hole and the Taub-NUT spacetime. The
main feature of this black hole is that it has 4 horizons like
event horizon, Cauchy horizon and two accelerating horizons. In
the non-extremal case, the surface area, entropy, surface gravity,
temperature, angular velocity, Komar energy and irreducible mass
on the event horizon and Cauchy horizon are presented for PD black
hole. The entropy product, temperature product, Komar energy
product and irreducible mass product are found for event horizon
and Cauchy horizon. Also their sums are also found for both
horizons. All these relations are found to be depend on mass of
the PD black hole and other parameters. So all the products are
not universal for PD black hole. The entropy and area bounds for
two horizons are investigated. Also we found the
Christodoulou-Ruffini mass for extremal PD black hole. Finally,
using first law of thermodynamics, we also found the Smarr
relation for
PD black hole.\\

\end{abstract}


\maketitle

\section{Introduction}
In 1981, Bekenstein \cite{Bek1} proposed the universal bound on
the entropy of a macroscopic object of maximal radius $R$ bearing
energy $E$ in the form $S\le \frac{2\pi ER}{\hbar}$. But the
derivation of the entropy bound was criticized by Unruh, Wald and
Pelath \cite{Wald1,Wald2,Wald3}. Bekenstein refused their
criticism and showed that buoyancy is so negligible such that it
does not spoil the entropy bound derivation \cite{Bek2,Bek3}. In
various occasions, this type of bound has been found in some
literatures by the same author \cite{Bek4,Bek5,Bek6}. In 1992,
Zaslavskii \cite{Zas} modified the entropy bound by incorporating
the charge of a black hole. After that Bekenstein and Mayo
\cite{Bek7} and Hod \cite{Hod1,Hod2} have obtained an upper
entropy bound for a charged Reissner-Nordstrom black hole, in the
form $S\le \frac{2\pi R}{\hbar}\left(E^{2}-\frac{e^{2}}{2R}
\right)$, where $e$ is the electric charge of the black hole.
Shimomura et al \cite{Shi} have also obtained the entropy bound
for charged black hole. Including the angular momentum, Linet
\cite{Linet1,Linet2} and Qiu et al \cite{Qiu} obtained the upper
bound of the entropy on the more general Kerr-Newman black hole.
Wand and Abdalla \cite{Wang} studied the entropy bound for a
spinning object falling into anti de Sitter (AdS) black holes
including $(3+1)$-dimensional Kerr-AdS black holes and
$(2+1)$-dimensional Banados-Teitelboim-Zanelli (BTZ) black holes.
Jing \cite{Jing} obtained the Cardy-Verline formula and the
entropy bounds in Kerr-Newman-AdS$_{4}/$dS$_{4}$ black hole.\\

The product of horizon areas or the entropy product of horizons of
black hole are very important in the study of black hole physics.
The two horizons of the black hole are namely inner/Cauchy horizon
$({\cal H}^{-})$ and outer/event horizon $({\cal H}^{+})$. Now it
is known that the Cauchy horizon $({\cal H}^{-})$ is an infinite
blue-shift surface, but the event horizon $({\cal H}^{+})$ is an
infinite red-shift surface \cite{Chandra}. For stationary axially
symmetric black holes, the entropy product of horizons are often
independent of the mass of the black hole
\cite{Cve,Viss,Tol,Chen,Cve1}. Such products depend on the charge
and angular momentum of the black hole. In some cases this
relation may be depend on the mass of the black hole
\cite{Cast,Vis,Xu1}. So the entropy sum and other thermodynamic
relations have been studied \cite{Prad,Prad1,Du,Wang1,Xu2}. In
some cases these relations may be independent of black hole mass
and some cases these depend on black hole mass. For regular
axisymmetric and stationary spacetime of an Einstein-Maxwell
system with surrounding matter has a regular Cauchy horizon
$({\cal H}^{-})$, which always occur inside the event horizon
$({\cal H}^{+})$ if and only if the angular momentum $J$ and
charge $Q$ of the black hole do not vanish simultaneously. In this
case, the Cauchy horizon $({\cal H}^{-})$ becomes singular and
tends to a curvature singularity when $J$ and $Q$ tend to zero
\cite{Ans1,Ans2,Ans3}. In Boyer-Lindquist coordinates, the
existence of Cauchy horizon describes that the stationary and
axisymmetric Einstein-Maxwell electro-vacuum equations are
hyperbolic in the interior vicinity of the event horizon $({\cal
H}^{+})$. The two horizons ${\cal H}^{+}$ and ${\cal H}^{-}$
describe the future and past boundary of this hyperbolic region.
If the Cauchy horizon exists i.e., if $J$ and $Q$ do not vanish
simultaneously, then the product of the entropy (area) of the
horizons ${\cal H}^{\pm}$ for the Kerr-Newman black hole is
independent of the mass of the black hole, but depends on the
angular momentum $J$ and the charge $Q$ explicitly \cite{Prad}.\\

Based on the entropy product and entropy sum, very recently, Xu et
al \cite{Xu} have obtained entropy (area) bound for event horizon
$({\cal H}^{+})$ and Cauchy horizon $({\cal H}^{-})$ for Kerr,
Kerr-Newman, Kerr-Newman in Gauss Bonnet gravity and Kerr-Taub-NUT
black holes. They have actually taken the Penrose-like inequality
for the upper area bound of event horizon $({\cal H}^{+})$. They
also found that the electric charge $Q$ diminishes the physical
bound of entropy (area) for event horizon $({\cal H}^{+})$, while
it enlarges that for Cauchy horizon $({\cal H}^{-})$; the angular
momentum $J$ enlarges them for Cauchy horizon $({\cal H}^{-})$,
while it does nothing with that for event horizon $({\cal
H}^{+})$; the NUT charge $n$ always enlarges them for both event
horizon $({\cal H}^{+})$ and Cauchy horizon $({\cal H}^{-})$. With
the ideas of their work, we now formulate the entropy bounds on
event horizon $({\cal H}^{+})$ and Cauchy horizon $({\cal H}^{-})$
for more general accelerating, rotating and charged
Plebanski-Demianski black hole. We also determine the entropy
product, entropy sum, sum of the angular velocities,
temperature-entropy relations, black hole mass and the entropy
bounds for event horizon $({\cal H}^{+})$ and Cauchy horizon
$({\cal H}^{-})$. Finally we discuss the main results of the work.

\section{Plebanski-Demianski Black Hole}

A large family of Einstein-Maxwell electro-vacuum solutions of
algebraic type $D$ was presented by Plebanski and Demianski
\cite{PD}. This includes the Kerr-Newman rotating black holes, the
Taub-NUT spacetime, the (anti-) de Sitter metric and also their
arbitrary combination. The resulting black holes may accelerate
due to conical singularities (such as cosmic strings). The general
form of the metric contains eight free parameters which
characterize the mass, charge (electric and magnetic), angular
momentum, rotation parameter, the NUT parameter, acceleration of
the sources and the cosmological constant
\cite{Griff1,Griff2,Griff3,Pod1,Pod2}. The new form of
accelerating, rotating and charged Plebanski-Demianski (PD) black
hole metric is given by \cite{Griff1}
\begin{eqnarray*}
ds^{2}=\frac{1}{\Omega^{2}}\left[-\frac{Q}{\rho^{2}}\left\{dt-\left(a~sin^{2}\theta+4l
sin^{2}\frac{\theta}{2}\right)d\phi
\right\}^{2}+\frac{\rho^{2}}{Q}~dr^{2} \right.
\end{eqnarray*}
\begin{equation}
\left. +\frac{P}{\rho^{2}}~\left\{adt-(r^{2}+(a+l)^{2})d\phi
\right\}^{2}+\frac{\rho^{2}}{P}~sin^{2}\theta d\theta^{2} \right]
\end{equation}
where $\Omega,~\rho^{2},~Q$ and $P$ are in the following:
\begin{equation}
\Omega=1-\frac{\alpha}{\omega}~(l+a~cos\theta)r~,
\end{equation}
\begin{equation}
\rho^{2}=r^{2}+(l+a~cos\theta)^{2}~,
\end{equation}
\begin{eqnarray*}
Q=\left[(\omega^{2}k+e^{2}+g^{2})(1+\frac{2\alpha
l}{\omega}~r)-2Mr+\frac{\omega^{2}k}{a^{2}-l^{2}}~r^{2} \right]
\end{eqnarray*}
\begin{equation}
\times
\left[1-\frac{\alpha(a+l)}{\omega}~r\right]\left[1+\frac{\alpha(a-l)}{\omega}~r\right]~,
\end{equation}
\begin{equation}
P=(1-a_{3}~cos\theta-a_{4}~cos^{2}\theta)sin^{2}\theta~,
\end{equation}
\begin{equation}
a_{3}=\frac{2\alpha
a}{\omega}~M-\frac{4\alpha^{2}al}{\omega^{2}}~(\omega^{2}k+e^{2}+g^{2})~,
\end{equation}
\begin{equation}
a_{4}=-\frac{\alpha^{2}a^{2}}{\omega^{2}}~(\omega^{2}k+e^{2}+g^{2})~,
\end{equation}
\begin{equation}
\left[\frac{\omega^{2}}{a^{2}-l^{2}}+3\alpha^{2}l^{2}\right]k=1+\frac{2\alpha
l}{\omega}~M-\frac{3\alpha^{2}l^{2}}{\omega^{2}}~(e^{2}+g^{2})
\end{equation}
Here, $e$ is the electric charge, $g$ is the magnetic charge, $a$
is the angular momentum $(=J/M)$, $l$ is the NUT parameter,
$\alpha$ is the acceleration and $\omega$ is the rotation
parameter. The PD black hole metric reduces to the following well
known black hole metrics as special cases: (i)
Kerr-Newman-Taub-NUT $(\alpha=g=0)$ \cite{Mill,Bini} (ii)
Kerr-Taub-NUT $(\alpha=e=g=0)$ \cite{Bini1} (iii) Taub-NUT
$(\alpha=a=e=g=0)$ \cite{Iwai} (iv) Kerr-Newman $(\alpha=l=g=0)$
\cite{Lee} (v) Kerr $(\alpha=l=e=g=0)$ \cite{Bard} (vi)
Riessner-Nordstrom $(\alpha=a=l=g=0)$ (vii) Schwarzschild
$(\alpha=a=l=e=g=0)$ and (viii) C-metric $(a=l=0)$ \cite{Hong1}.\\

Horizons of the black hole found when $\frac{1}{g_{rr}}=0$, i.e.,
$Q=0$. So the horizons are located at
\begin{eqnarray*}
r_{\pm}=\frac{a^{2}-l^{2}}{\omega^{2}k}\left[\left(M-\frac{\alpha
l}{\omega}~(\omega^{2}k+e^{2}+g^{2}) \right)
\right.~~~~~~~~~~~~~~~~~~~~~~
\end{eqnarray*}
\begin{equation}
\left. \pm  \sqrt{\left(M-\frac{\alpha
l}{\omega}~(\omega^{2}k+e^{2}+g^{2}) \right)^{2}
-\frac{\omega^{2}k}{a^{2}-l^{2}}~(\omega^{2}k+e^{2}+g^{2})}~
 \right]
\end{equation}
with $|a|>|l|$. Here $r_{+}$ represents the outer/event horizon
$({\cal H}^{+})$ and $r_{-}$ represents the inner/Cauchy horizon
$({\cal H}^{-})$. Also $r=\pm \frac{\omega}{\alpha(a \pm l)}$ are
known as the accelerating horizons. While $g_{tt}=0$, i.e.,
$Q=a^{2}P$ describes the ergosphere. Here we are interested to
study only the event horizon ${\cal H}^{+}$ and Cauchy horizon
${\cal H}^{-}$. Now define the product and addition of the roots
$r_{+}$ and $r_{-}$ by $A=r_{+}r_{-}$ and $B=r_{+}+r_{-}$, where
\begin{equation}
A=\frac{(a^{2}-l^{2})}{\omega^{2}k}(\omega^{2}k+e^{2}+g^{2})
\end{equation}
and
\begin{equation}
B=\frac{2(a^{2}-l^{2})}{\omega^{2}k}\left(M-\frac{\alpha
l}{\omega}~(\omega^{2}k+e^{2}+g^{2}) \right)
\end{equation}
In order to avoid naked singularity, $r_{\pm}$ must be real. So we
have the restriction:
\begin{equation}\label{12}
\left(M-\frac{\alpha l}{\omega}~(\omega^{2}k+e^{2}+g^{2})
\right)^{2} \ge
\frac{\omega^{2}k}{a^{2}-l^{2}}~(\omega^{2}k+e^{2}+g^{2})
\end{equation}
with
\begin{equation}
k\ge 0 ~~ and ~~M\ge \frac{\alpha
l}{\omega}~(\omega^{2}k+e^{2}+g^{2})
\end{equation}
If we take the equality of the expression (\ref{12}), we must have
$r_{+}=r_{-}$. In this situation, the two horizons will coincide
and the black hole will be called the {\it extremal black hole}.
Now we assume that the black hole is non-extremal (i.e., if there
exists a trapped surface interior of the outer horizon) so
$r_{+}>r_{-}$. Now the surface area of the PD black hole on ${\cal
H}^{\pm}$ is obtained by \cite{Prad}
\begin{equation}
{\cal A}_{\pm}=\int\int\sqrt{g_{\theta\theta}g_{\phi\phi}}~d\theta
d\phi=\frac{4\pi\omega^{2}(r_{\pm}^{2}+(a+l)^{2})}{(\omega-l\alpha
r_{\pm})^{2}-a^{2}\alpha^{2}r^{2}_{\pm}}
\end{equation}
So the entropies on the horizons ${\cal H}^{\pm}$ are
\begin{equation}
S_{\pm}=\frac{{\cal
A}_{\pm}}{4}=\frac{\pi\omega^{2}(r_{\pm}^{2}+(a+l)^{2})}{(\omega-l\alpha
r_{\pm})^{2}-a^{2}\alpha^{2}r^{2}_{\pm}}
\end{equation}
Now the surface gravities on the horizons ${\cal H}^{\pm}$ can be
obtained by
\begin{equation}
\kappa_{\pm}=\frac{r_{\pm}-r_{\mp}}{2(r_{\pm}^{2}+(a+l)^{2})}
\end{equation}
Since Chen et al \cite{Chen} defined
$T_{-}=-T_{+}|_{r_{+}\leftrightarrow r_{-}}$ so the temperatures
on the horizons ${\cal H}^{\pm}$ are given by
\begin{equation}
T_{\pm}=\frac{|\kappa_{\pm}|}{2\pi}=\pm
\frac{r_{\pm}-r_{\mp}}{4\pi(r_{\pm}^{2}+(a+l)^{2})}
\end{equation}
Since $r_{+}>r_{-}$, so $T_{+}<T_{-}$. Hence we may say that the
Cauchy horizon is hotter than the event horizon. The product of
the temperatures on the both horizons ${\cal H}^{\pm}$ yields to
be
\begin{equation}
T_{+}T_{-}=\frac{B^{2}-4A}{16\pi^{2}\left[A^{2}+(a+l)^{2}(B^{2}-2A)+(a+l)^{4}\right]}
\end{equation}
Curir \cite{Curir} introduced the `area sum' and `entropy sum' of
Kerr BH for the interpretation of the spin entropy of the area of
the inner horizon. Now for PD black hole the product, addition and
subtraction of two horizons (${\cal H}^{\pm}$) entropies are
obtained as in the following:
\begin{equation}
S_{+}S_{-}=\frac{F}{G}~,
\end{equation}
\begin{equation}
S_{+}+S_{-}=\frac{H}{G}~,
\end{equation}
and
\begin{equation}
S_{+}-S_{-}=\frac{I}{G}
\end{equation}
where
\begin{equation}
F=\pi^{2}\omega^{4}\left[A^{2}+(a+l)^{4}+(a+l)^{2}(B^{2}-2A)
\right]~,
\end{equation}
\begin{eqnarray*}
G=(a^{2}-l^{2})^{2}\alpha^{4}A^{2}+2l\alpha^{3}\omega(a^{2}-l^{2})AB~~~~~~~~~~~~~~~~~~~~~~~~~~~~~~~~~~~~~
\end{eqnarray*}
\begin{equation}
-\omega^{2}\alpha^{2}(a^{2}-l^{2})(B^{2}-2A)+4l^{2}\alpha^{2}\omega^{2}A-2l\alpha\omega^{3}B+\omega^{4}~,
\end{equation}
\begin{eqnarray*}
H=\pi\omega^{2}\left[2\alpha^{2}(l^{2}-a^{2})A^{2}-2l\alpha\omega
AB \right.
\end{eqnarray*}
\begin{eqnarray*}
+\{(a+l)^{2}(l^{2}-a^{2})\alpha^{2}+\omega^{2} \} (B^{2}-2A)
\end{eqnarray*}
\begin{equation}
\left.  -2l\alpha\omega(a+l)^{2}B+2\omega^{2}(a+l)^{2} \right]~,
\end{equation}
\begin{eqnarray*}
I=\pi\omega^{2}(r_{+}-r_{-})\left[\{\omega^{2}+(a^{2}-l^{2})(a+l)^{2}\alpha^{2}
\}(r_{+}-r_{-}) \right.
\end{eqnarray*}
\begin{equation}
\left. +2l\alpha\omega\{(a+l)^{2}-A \} \right]~.
\end{equation}
From the above relations we may obtain the sum of entropy inverse
\begin{equation}
\frac{1}{S_{+}}+\frac{1}{S_{-}}=\frac{H}{F}
\end{equation}
The angular velocities on the horizons ${\cal H}^{\pm}$ are
obtained by
\begin{equation}
\Omega_{\pm}=-\frac{g_{t\phi}}{g_{\phi\phi}}=\frac{a}{r_{\pm}^{2}+(a+l)^{2}}
\end{equation}
The sum of the angular velocities on ${\cal H}^{\pm}$ are
\begin{equation}
\Omega_{+}+\Omega_{-}=\frac{a[B^{2}-2A+2(a+l)^{2}]}{A^{2}+(a+l)^{2}(B^{2}-2A)+(a+l)^{4}}
\end{equation}
The Komar \cite{Komar} energy for ${\cal H}^{\pm}$ is given by
\begin{equation}
E_{\pm}=2T_{\pm}S_{\pm}=\pm \frac{\omega^{2}}{2}~
\frac{r_{\pm}-r_{\mp} }{(\omega-l\alpha
r_{\pm})^{2}-a^{2}\alpha^{2}r^{2}_{\pm}}
\end{equation}
The product of Komar energies on the two horizons $({\cal
H}^{\pm})$ are obtained by
\begin{equation}
E_{+}E_{-}=\frac{\omega^{4}}{4G}~(B^{2}-4A)
\end{equation}

Penrose et al \cite{Penrose} shown that when a black hole (eg.
Kerr black hole) is undergoing any transformations, the surface
area of the horizon increases, which is known as Penrose process.
Independently, Christodoulou \cite{Christ} also shown that the
mass of the black hole unchanged by any process, which is known as
Christodoulou's ``irreducible mass" denoted by ${\cal M}$. But
most of the processes, it was seen that ${\cal M}$ increases and
during reversible process this quantity also does not change. From
this result, it was conclude that there exists a relation between
the area and irreducible mass. The irreducible mass ${\cal
M}_{\pm}$ for the horizons ${\cal H}^{\pm}$ can be defined by
\cite{Prad}
\begin{equation}
{\cal M}_{\pm}=\sqrt{\frac{{\cal A}_{\pm}}{16\pi}}
\end{equation}
Now the product of the irreducible mass for PD black hole is
obtained by
\begin{equation}
{\cal M}_{+}{\cal M}_{-}=\frac{1}{4\pi}\sqrt{\frac{F}{G}}
\end{equation}

From all the products of the thermodynamic quantities for ${\cal
H}^{\pm}$, we observe that these products completely depend on the
mass of PD black hole $M$ and other parameters. So we we may
conclude that the products are not universal, while Pradhan
\cite{Prad1} claimed that for charged rotating Kerr-Newman black
hole, the thermodynamic quantities are not also universal except
the area product and entropy product (because they do not depend
of black hole mass). The relation of temperatures and entropies on
the two horizons ${\cal H}^{\pm}$ are \cite{Curir}
\begin{equation}
T_{+}S_{+}-T_{-}S_{-}=\frac{\omega^{2}}{4G}~(r_{+}-r_{-})[(a^{2}-l^{2})\alpha^{2}B+2l\alpha\omega]
\end{equation}
If there is no acceleration of the black hole i.e., $\alpha=0$, we
obtain $T_{+}S_{+}=T_{-}S_{-}$. This reads that the central charge
for two-horizons black hole are same. But for accelerating black
hole, $T_{+}S_{+}\ne T_{-}S_{-}$. Since $r_{+}>r_{-}$, so we have
${\cal A}_{+}>{\cal A}_{-}\ge 0$. Now from the entropy product, we
find \cite{Xu}
\begin{equation}
S_{+}\ge \sqrt{S_{+}S_{-}}=\sqrt{\frac{F}{G}}\ge S_{-}\ge 0.
\end{equation}
Also from the entropy sum \cite{Xu}, we obtain
\begin{equation}
\frac{H}{G}=S_{+}+S_{-}\ge S_{+}\ge
\frac{S_{+}+S_{-}}{2}=\frac{H}{2G}\ge S_{-}
\end{equation}
Thus the entropy bounds for event horizon (${\cal H}^{+}$) and
Cauchy horizon (${\cal H}^{-}$) of PD black hole are
\begin{equation}
S_{+} \in \left[\frac{H}{2G}~,~\frac{H}{G} \right]~,~~~S_{-} \in
\left[0~,~\sqrt{\frac{F}{G}}  \right]
\end{equation}
So the area bounds for event horizon (${\cal H}^{+}$) and Cauchy
horizon (${\cal H}^{-}$) of PD black hole are
\begin{equation}
{\cal A}_{+} \in \left[\frac{2H}{G}~,~\frac{4H}{G}
\right]~,~~~{\cal A}_{-} \in \left[0~,~4\sqrt{\frac{F}{G}} \right]
\end{equation}
and hence the bounds of the irreducible mass for event horizon
(${\cal H}^{+}$) and Cauchy horizon (${\cal H}^{-}$) of PD black
hole are obtained as
\begin{equation}
{\cal M}_{+} \in \left[\sqrt{\frac{H}{8\pi
G}}~,~\sqrt{\frac{H}{4\pi G}} \right]~,~~~{\cal M}_{-} \in
\left[0~,~\frac{1}{\sqrt{4\pi}}\left(\frac{F}{G}
\right)^{\frac{1}{4}} \right]
\end{equation}
From the above bounds for horizons entropy, area and irreducible
mass, we understand that the lower and upper bounds completely
depend on all the parameters like PD black hole mass $M$, angular
momentum $a$, electric charge $e$, magnetic charge $g$, rotation
parameter $\omega$, NUT parameter $l$ and the acceleration
parameter $\alpha$. In particular, for extremal PD black hole,
$r_{+}=r_{-}$, we obtain ${\cal A}_{+}={\cal A}_{-}$,
$S_{+}=S_{-}$, $\Omega_{+}=\Omega_{-}$, ${\cal M}_{+}={\cal
M}_{-}$, $\kappa_{+}=\kappa_{-}=0$, $T_{+}=T_{-}=0$ and
$E_{+}=E_{-}=0$. So surface gravity, temperature and Komar energy
vanish on the horizon of the PD extremal black hole. For extremal
PD black hole, we may obtain the area $A=\frac{H}{8G}$~, entropy
$S=\frac{H}{2G}$ and angular velocity
$\Omega=\frac{a}{A+(a+l)^{2}}$ with the conditions $I=0$,
$B^{2}=4A$ and $H^{2}=4FG$. And hence the Christodoulou-Ruffini
\cite{CR} mass  for extremal PD black hole will be
\begin{eqnarray*}
{\cal M}_{CR}=
\left[l^{3}\alpha^{3}\omega(a^{2}-l^{2})((a^{2}-l^{2})-2(e^{2}+g^{2}))
\right.
\end{eqnarray*}
\begin{eqnarray*}
+l\alpha\omega^{3}(3(a^{2}-l^{2})+2(e^{2}+g^{2}))+\omega(3l^{2}\alpha^{2}(a^{2}-l^{2})+\omega^{2})
\end{eqnarray*}
\begin{eqnarray*}
\left. \times
\sqrt{\omega^{2}(a^{2}-l^{2}+e^{2}+g^{2})-l^{2}\alpha^{2}(a^{2}-l^{2})(e^{2}+g^{2})}
\right]
\end{eqnarray*}
\begin{equation}
\times\left[\omega^{2}-l^{2}\alpha^{2}(a^{2}-l^{2})\right]^{-2}
\end{equation}
with the condition
\begin{equation}
\omega^{2}(a^{2}-l^{2}+e^{2}+g^{2})\ge
l^{2}\alpha^{2}(a^{2}-l^{2})(e^{2}+g^{2}).
\end{equation}
In particular, for $\alpha=g=0$, we can recover the
Christodoulou-Ruffini mass for extremal Kerr-Newman-Taub-NUT black
hole \cite{Prad} i.e., ${\cal M}_{CR}=\sqrt{a^{2}+e^{2}-l^{2}}$.\\

Finally, we demonstrate the first law of thermodynamics of event
horizon (${\cal H}^{+}$) and Cauchy horizon (${\cal H}^{-}$) for
PD black hole. The first law of thermodynamics are given by
\cite{Xu}
\begin{equation}
dM=\pm T_{\pm}dS_{\pm}+\Omega_{\pm}dJ+{\cal E}_{\pm}de+{\cal
G}_{\pm}dg+\Phi_{\pm} dl
\end{equation}
where ${\cal E}_{\pm}$, ${\cal G}_{\pm}$ and $\Phi_{\pm}$ are the
electromagnetic potentials for electric charge, magnetic charge
and NUT charge respectively defined by \cite{Xu}
\begin{eqnarray*}
{\cal E}_{\pm}=\left(\frac{\partial M}{\partial
e}\right)_{S_{\pm},J,g,l}~,~{\cal G}_{\pm}=\left(\frac{\partial
M}{\partial g}\right)_{S_{\pm},J,e,l}~,
\end{eqnarray*}
\begin{equation}
~\Phi_{\pm}=\left(\frac{\partial M}{\partial
l}\right)_{S_{\pm},J,e,g}
\end{equation}
Here $M$ scales as [length]$^{1}$, $S_{\pm}$ scales as
[length]$^{2}$, $J$ scales as [length]$^{2}$, $e$ scales as
[length]$^{1}$, $g$ scales as [length]$^{1}$ and $l$ scales as
[length]$^{1}$ \cite{Xu}. So we may obtain
\begin{equation}
M=2(\pm T_{\pm}S_{\pm}+\Omega_{\pm}J)+{\cal E}_{\pm}e+{\cal
G}_{\pm}g+\Phi_{\pm} l
\end{equation}
which behave as the Smarr relation \cite{Smarr,Smarr1} of PD black
hole for event and Cauchy horizons ${\cal H}^{\pm}$.

\section{Discussions}

In this work, first we have reviewed the most general black hole
i.e., the accelerating, rotating and charged Plebanski-Demianski
(PD) black hole, which includes the Kerr-Newman rotating black
hole and the Taub-NUT spacetime. Neglecting some parameters
involved in the PD black hole metric, we can recover the
Kerr-Newmann-Taub-NUT black hole, Kerr-Taub-NUT black hole,
Taub-NUT black hole, Kerr-Newmann black hole, Kerr black hole,
Riessner-Nordstrom black hole, Schwarzschild black hole and
C-metric. The main feature of this PD black hole is that it has 4
horizons like event horizon (${\cal H}^{+}$), Cauchy horizon
(${\cal H}^{-}$) and two accelerating horizons which are located
at $r_{+}$, $r_{-}$ and $r=\pm\frac{\omega}{\alpha(a\pm l)}$
respectively. Since the event horizon (${\cal H}^{+}$) forms
outside the Cauchy horizon (${\cal H}^{-}$), so in the
non-extremal case (i.e., $r_{+}> r_{-}$), the surface area,
entropy, surface gravity, temperature, angular velocity, Komar
energy and irreducible mass on the event horizon and Cauchy
horizon are presented for PD black hole. Since $r_{+}>r_{-}$, so
$T_{+}<T_{-}$ i.e., we may say that the Cauchy horizon is hotter
than the event horizon. But for extremal case (two horizons are
identical i.e., $r_{+}=r_{-}$), we have found that the surface
gravity, temperature and Komar energy vanish on the horizon. The
entropy product, temperature product, Komar energy product and
irreducible mass product are found for event horizon and Cauchy
horizon. Also their sums are also found for both horizons. All
these relations are found to be depend on mass of the PD black
hole and other parameters involved. So all the products are said
to be not universal for PD black hole. Also from the relations of
temperature and entropy of the horizons ${\cal H}^{\pm}$, we found
that $T_{+}S_{+}\ne T_{-}S_{-}$. If there is no acceleration of
the black hole i.e., $\alpha=0$, we can recover
$T_{+}S_{+}=T_{-}S_{-}$. This reads that for non-accelerating
black hole, the central charge for two-horizons black hole are
same. Since $r_{+}>r_{-}$, so we have ${\cal A}_{+}>{\cal
A}_{-}\ge 0$ and from the entropy product and sum, we found the
bounds of the entropy for ${\cal H}^{\pm}$ as $S_{+} \in
\left[\frac{H}{2G}~,~\frac{H}{G} \right],~S_{-} \in
\left[0~,~\sqrt{\frac{F}{G}} \right]$. Also we found the bounds of
area for ${\cal H}^{\pm}$. All these bounds are completely depend
on the 7 parameters $M,~e,~g,~\omega,~\alpha,~a,~l$. For extremal
(two horizons are identical) black hole, we found the area
$A=\frac{H}{8G}$~, entropy $S=\frac{H}{2G}$ and angular velocity
$\Omega=\frac{a}{A+(a+l)^{2}}$ with the conditions $I=0$,
$B^{2}=4A$ and $H^{2}=4FG$. Also we found the
Christodoulou-Ruffini mass for extremal PD black hole. Finally,
using first law of thermodynamics, we also found the Smarr
relation for PD black hole.\\

{\bf Acknowledgement:} The author is thankful to
IUCAA, Pune, India for warm hospitality where the work was carried out.\\

\end{document}